\documentclass[12pt]{article}

\usepackage{amssymb,bbm}
\usepackage{amsfonts}
\usepackage{latexsym}
\usepackage{amsmath}
\usepackage{amsthm}
\usepackage{geometry} 
\usepackage{graphicx}
\usepackage{verbatim}
\usepackage{bm} 

\newtheorem{thm}{Theorem}[section]
\newtheorem{prop}[thm]{Proposition}
\newtheorem{lemma}[thm]{Lemma}

\newtheorem{defi}[thm]{Definition}
\newcommand{\rank}{\text{rank}}
\newcommand{\prf}{\underline{Proof:}\ }
\newcommand{\finprf}{\null \hfill {\rule{5pt}{5pt}}\\[2.1ex]\indent}

\newcommand{\II}{{\mathbb I}}

\def\lddots{\mathinner{\mkern1mu\raise1pt\hbox{.}\mkern2mu  
\raise4pt\hbox{.}\mkern2mu\raise7pt\vbox{\kern7pt\hbox{.}}\mkern1mu}}
\makeatletter
\def\numberbysection{\@addtoreset{equation}{section}
 \def\theequation{\thesection.\arabic{equation}}}
\makeatother  
  
\numberbysection

\newcommand{\be}{\begin{eqnarray}}  
\newcommand{\ee}{\end{eqnarray}} 
  
\def\ds{\displaystyle}

\def\bb{\mathbbm}
\def\C{\bb C}

 \title{\bf Temperley-Lieb $R$-matrices from \\ generalized Hadamard matrices}
\author{ \textsf{Jean Avan$^a$}
\textsf{, Tiago Fonseca$^b$}
\textsf{, Luc Frappat$^b$}
\\
\textsf{ Petr Kulish$^c$}
\textsf{, Eric Ragoucy$^b$}
\textsf{ and Genevi\`eve Rollet$^a$}
\\\\
\textit{$^{a}$ Laboratoire de Physique Th\'eorique et Mod\'elisation (CNRS UMR 8089),} \\
\textit{Universit\'e de Cergy-Pontoise, F-95302 Cergy-Pontoise, France} \\
\\
\textit{$^{b}$ LAPTh, CNRS and Universit\'e de Savoie,} \\
\textit{9 Chemin de Bellevue, BP 110, F-74941 Annecy le Vieux Cedex} \\
\\
\textit{$^c$
St. Petersburg Department of Steklov Mathematical Institute} \\
\textit{Fontanka 27, 191023, St. Petersburg, Russia} }
\date{}

\footnotetext[1]{Emails: {\tt avan@u-cergy.fr}, 
{\tt tiago.dinis@lapth.cnrs.fr},
{\tt frappat@lapth.cnrs.fr},
{\tt kulish@pdmi.ras.ru},
{\tt ragoucy@lapth.cnrs.fr},
{\tt rollet@u-cergy.fr}}

\begin{document}

\maketitle
\thispagestyle{empty}
\abstract{New sets of rank $n$-representations of Temperley-Lieb algebra $TL_N(q)$ are constructed. They are characterized by 
two matrices obeying a generalization of the complex Hadamard property. Partial classifications for the two matrices are given, 
in particular when they reduce to Fourier or Butson matrices.

\bigskip\bigskip

\rightline{LAPTH-029/13\qquad\qquad}}

\clearpage 
\newpage

\section{Introduction}

The Temperley-Lieb algebra (hereafter denoted TL) $TL_N(q)$~\cite{TL,Ba}, has been used extensively as a powerful algebraic tool in
the construction and derivation of quantum integrable models of great interest in statistical mechanics and solid state physics 
(see e.g. \cite{Ba,PM}).
Special representations of the TL algebra where the generators are copies of a single endomorphism
acting on a tensor product $V \otimes V$, $V$ being an $n$-dimensional vector space, 
give rise to constant solutions $R$ of the Yang-Baxter equation. Yang-Baxterization procedures 
are then systematically available (see e.g. \cite{AKR}).
From such Yang-Baxterized $R$-matrices one then may in particular
construct integrable quantum spin chains~\cite{PPK1} on the space of states
${\cal H}=\left(\mathbb{C}^n\right)^{\otimes N}$ for any integer $n$. 
These spin chains are very similar to the spin $1/2$ $XXZ$-model.

Specific representations of TL algebra were introduced in e.g. \cite{PPK1,KMN}: they are para\-me\-tri\-zed by a single
bivector $b$ yielding a rank-1 projector on $V \otimes V$. The Temperley-Lieb parameter $q$ to be defined hereafter
was identified by $q + \frac{1}{q} \equiv -tr (b b^t)$. A classification of solutions to the reflection
equation associated to the derived $R$-matrix was proposed in \cite{AKR}, aiming at building quantum
integrable open spin chains.

An extension of these representations involving $n$ such bivectors was proposed in \cite{Chin1,Chin2} as relevant
in the context of entanglement and quantum computing. The TL parameter or ``loop index'' is then identified by
$q + \frac{1}{q} \equiv \sqrt n$. The matrices, originally parametrized by $n$ bivectors, 
were naturally written as $n^2 \times n^2$ matrices as in e.g. the $n=3$ case of \cite{Chin2}:
\[
 U^{(I\!I)} = 
\begin{pmatrix}
  1 & 0 & 0 & 0 & 0 & \omega & 0 & 1 & 0\\
  0 & 1 & 0 & \omega & 0 & 0 & 0& 0 & \omega\\
  0 & 0 & 1 & 0 & 1 & 0  & 1 & 0 & 0\\
  0 & \omega^2 & 0 & 1 & 0 & 0  & 0 & 0 & 1\\
  0 & 0 & 1 & 0 & 1 & 0 & 1 & 0 & 0\\
  \omega^2 & 0 & 0 & 0 & 0 & 1 & 0 & \omega^2 & 0\\
  0 & 0 & 1 & 0 & 1 & 0 & 1 & 0 & 0\\
  1 & 0 & 0 & 0 & 0 & \omega & 0 & 1 & 0\\
  0 & \omega^2 & 0 & 1 & 0 & 0  & 0 & 0 & 1
 \end{pmatrix}
\]
where $\omega^2+\omega+1 =0$. It turns out (see below) that these new TL representations can more appropriately be 
reformulated in terms of a sum of $n^2$ ordinary tensor products of two $n\times n$ matrices, namely:
\be
T_i = \sum_{a,b=1}^n \II^{\otimes(i-1)}\otimes e_{ab} \otimes M^{n_{a} -n_{b}}\otimes \II^{\otimes(N-i-1)}\,,\quad i=1,...,N
\label{form}
\ee
where $e_{ab}$ denotes the canonical form of the generators of $n \times n$
matrices, $M$ is a single invertible $n \times n$ matrix and $n_{a}$ are integers.

Precisely the representation $ U^{(I\!I)}$ in \cite{Chin2} takes the form \eqref{form}
with :
\[
 M=
\begin{pmatrix}
0 & 1 & 0\\
0 & 0 & \omega\\
1 & 0 & 0
 \end{pmatrix}
\]
In an explicit way $U^{(I\!I)}$ reads:
\be
 U^{(I\!I)} = 
\begin{pmatrix}
M^0 & M^2 & M\\
M^{-2} & M^0 & M^{-1}\\
M^{-1} & M & M^0
 \end{pmatrix}
\label{U_2}
\ee
As usual in such representations the  $i$-th generator $T_i$ of TL acts non trivially only on
the two copies of $V$ labeled by resp. $i$ and $i+1$ in the full tensorized representation 
space $\displaystyle \bigotimes_{k=1}^{N} V_{(k)}$. The $R$ matrix deduced from
such an object is simply $R_{i,i+1} = \Pi_{i,i+1} (q\, \II \otimes \II + T_i)$ where $\Pi_{i,j}$ generically denotes the permutation operator on tensorized spaces $V_i \otimes V_j$, and $\II$ the identity. 
In our study, the $M$ matrix will be restricted to be diagonalizable. Jordan-reducible matrices shall be considered elsewhere.

This provides us with an interesting example of rank-$n$ realizations of the TL algebra
and motivates our current investigation of such generic realizations. The study of associated
scalar reflection matrices can be achieved on lines following \cite{AKR} but will be left for
another paper. One may expect that the new solutions which we propose here may be of interest,
again in the description of quantum entanglement effects, or more canonically as building blocks for
closed or open spin-chain like models after Yang-Baxterization. We shall come back to this in our conclusion.

The presentation runs as follows. In Section 2 we prepare the necessary notations, introduce 
precisely the Temperley-Lieb algebra and the rank-$n$ Ansatz which we use.
We then derive the relevant equations to be solved for a complete resolution based on this Ansatz \eqref{form}.

In Section 3 we separate these equations into a polynomial equation (denoted \textit{Master Polynomial} equation) controlling the 
eigenvalues of $M$ and a matrix equation controlling the eigenvectors of $M$.
Remarkably both sets are characterized by 
$n \times n$ matrices obeying an extension which we define (\textit{General Hadamard Condition} or GHC)
 of the Complex Hadamard property \cite{Hadcom,Pol}.
We then discuss the explicit classification of eigenvalues and eigenvectors based on these relations. 
The eigenvalues are encapsulated into a \textit{Master Matrix} obeying the general Hadamard condition. It 
however involves delicate issues  not yet fully clarified, since the Hadamard condition is here necessary but not sufficient. 
The eigenvectors by contrast are entirely determined by the choice of an arbitrary generalized Hadamard
matrix once the Master Matrix is known.

A partial set of solutions to the Complex Hadamard condition and its generalization is given in Section 4.
The representation $U^{(I)}$ in \cite{Chin1} is a simple example of a slightly more general set of objects
which is discussed in Section $5$. Finally we give some conclusions and perspectives.

\section{General properties and equations}

Let us first recall the general context of our discussion and obtain the equations to be solved to
get at least a partial classification of the solutions.

\subsection{Hecke and Temperley-Lieb Algebras}
The braid group  $\mathcal{B}_N$ is generated by $(N-1)$ generators 
$\check{R}_j$, $j=1,2,\ldots,N-1$, their inverses $\check{R}_j^{-1}$ and the relations (see~\cite{CP}):
\begin{eqnarray}
\label{BG}
\check{R}_j\check{R}_k  \check{R}_j  =  \check{R}_k \check{R}_j  \check{R}_k , \ \mbox{for} \ |j-k| =1 \quad \mbox{and}\quad
\check{R}_j\check{R}_k  =  \check{R}_k \check{R}_j , \ \mbox{for} \ |j-k| > 1.
\end{eqnarray}

Both Hecke algebra $H_N(q)$ and Temperley-Lieb algebra $TL_N(q)$ are quotients 
of the group algebra of $\mathcal{B}_N$:

\textbf{The Hecke algebra} $H_N(q)$ is obtained by adding to these 
relations the following constraints  obeyed by each generator $\check{R}_j$ ($q$-deformation of the symmetric group):
\be
\label{cheqR}
\left( \check{R}_j - q\, \II \right) \left( \check{R_j} + 1/q \, \II \right) = 0. 
\ee
where $\II$ denotes the identity in the Hecke algebra.
Equation \eqref{cheqR} is equivalent to  write $\check{R_j}$ in term of some idempotent $X_j$, namely:
\be
\label{RtoX}
\check{R_j} =  q\, \II + X_j
\ee
with 
\be
\label{Xproj}
X_j^2 =  - \left( q + \frac{1}{q} \right) X_j.
\ee
The braid group relations \eqref{BG} read in terms of the idempotents $X_j$ and $X_k$ such that $|j-k|=1$:
\begin{equation}
\label{BGX}
X_jX_kX_j - X_j = X_kX_jX_k - X_k . 
\end{equation}
\textbf{The TL algebra} $TL_N(q)$ is obtained as the quotient algebra of the Hecke algebra $H_N(q)$ by 
the set of equations requiring that each side of 
\eqref{BGX} be  zero.  
To sum up, $TL_N(q)$  is defined by the generators $X_j$, $j=1,2,\ldots, N-1$ and their relations:
\begin{eqnarray}
&&X_j^2 =  - \nu (q) X_j, \label{TLrel0} \\
&&X_jX_kX_j = X_j , \quad | j - k | = 1,\label{TLrel1} \\
&&X_jX_k = X_kX_j , \quad | j - k | > 1
\label{TLrel2}
\end{eqnarray}
with $\nu(q)=q+1/q$.

\medskip

In connection with integrable spin systems we will be interested in representations of $TL_N(q)$ 
on the tensor product space ${\cal H} = \left(\mathbb{C}^n\right)^{\otimes N}$. 
The $\check{R_j}$ generators are now represented in terms of endomorphisms on ${\cal H}$ 
acting non trivially on a pair $(j,j+1)$ of adjacent spaces $V$. These endomorphisms are 
self-explanatorily denoted as $\check{R}_{j,j+1}$. Conditions \eqref{BG} are then represented as the 
braided Yang-Baxter equation:
\be
\label{ConstBrYB}
\check{R}_{12}\ \check{R}_{23}\ \check{R}_{12} =  \check{R}_{23}\ \check{R}_{12}\ \check{R}_{23}.
\ee
The $R$-matrix is immediately defined from this representation of the braid group generators  
by $R_{j\, j+1}=\Pi_{j\, j+1} \check{R}_{j\, j+1}$, with $\Pi( v \otimes v' )
=v'\otimes v $ for any pair of vectors of $\C^n$. The indexation $(j\, j+1)$ of $\Pi$ is again self-explanatory.
The notation $R_{j\, j+1}$ is then straightforwardly extended to define general endomorphisms $R_{ij}$ of $\mathcal{H}$
labeled by any non-adjacent pair of ``site indices'' $(i,j)$, using the time-honored notation \cite{FaTa} for such
elements of $End({\cal H})$ with indices labelling the spaces.

Equation \eqref{ConstBrYB} then immediately becomes the Yang-Baxter equation for $R$:
\be
\label{ConstYB}
R_{12}\ R_{13}\ R_{23} = R_{23} \ R_{13} \ R_{12}. 
\ee
Of course any matrix realization of the YB algebra \eqref{ConstYB} can be gauged to another matrix realization
by the conjugation $R^g_{ij} \equiv g \otimes g\ R_{ij}\ g^{-1} \otimes g^{-1}$ where $g$ is any invertible
$n \times n$ matrix. This gauging freedom, naturally also valid for the considered TL representations,
 will be used in our reformulation of the Ansatz for TL representations.

Let us finally formulate the Yang-Baxterization procedure of these $R$-matrices. In fact the Yang-Baxterization procedure
is already valid at the stage of abstract Hecke algebra generators. Indeed if one defines
the spectral parameter-dependent $R$-matrix as \cite{Jones}
\be
\check{R}_{j}(u)= u \check{R}_{j} - \ds{\frac1u} \check{R}_{j}^{-1} = ( u - \ds{\frac1u}) \check{R}_{j} + \frac{\omega(q)}{u}\,\II;
\qquad \omega(q) = q - \frac{1}{q}
\label{baxtRm}
\ee
one sees that it obeys the cubic equation in braid group form
with multiplicative spectral parameter $u$ (additive spectral parameter is of course obtained as $u \equiv e^{\lambda}$):
\be
\check{R}_j(u)\check{R}_k (uw)  \check{R}_j(w)  =  \check{R}_k(w) \check{R}_j(uw)  \check{R}_k(u) , \quad \mbox{for} \ |j-k| =1.
\label{baxtH}
\ee
Now once the 
generators $\check R$ of the Hecke algebra $H_N(q)$ itself have been represented as $R$-matrices acting on some tensor product
of two finite-dimensional vector spaces, this procedure will
immediately (see \cite{AKR}) give rise to solutions of the non-constant braided Yang-Baxter equation 
with multiplicative spectral parameters:
\be
 \check{R}_{12}(u) \check{R}_{23}(uw) \check{R}_{12}(w)= \check{R}_{23}(w) \check{R}_{12}(uw) \check{R}_{23}(u).
\label{YBSP}
\ee 

\subsection{The rank-$n$ Ansatz and the master equation}

The initial construction of a rank-1 TL representation was proposed in \cite{PPK1}. The $U$ generators are represented by copies
of a projector onto a single bivector in $V_i \otimes V_{i+1}$. Extensions of this construction were then defined 
in \cite{Chin1,Chin2} as sum of $n$ projectors on $n$ bivectors $A^{(k)}$ along $n$ bivectors $B^{(k)}$ together
with consistency conditions. Exact solutions however were only derived for dimensions $n=2,3$ whereas we shall here 
derive general solutions for any $n$. 

A remark regarding the normalization is of order here. The representations in \cite{Chin1, Chin2} realize the exact
formulation of the TL algebra \eqref{TLrel0}-\eqref{TLrel2} with the one-loop factor $\nu(q) = \sqrt n$. To eliminate these 
awkward $\sqrt n$ factors we redefine the generators $U_i$
by an overall multiplication by $\sqrt n$. The one-loop equation \eqref{TLrel0} then gets a factor $n$ and the 
equation \eqref{TLrel1} acquires a factor $n$ on the r.h.s. It is these renormalized equations that we shall study from now on.

\begin{lemma}
Let $M$ be an invertible diagonalizable $n\times n$ matrix:
$M = P \Lambda P^{-1}$, where $\Lambda$ is  diagonal, $\Lambda=\mbox{diag}\big(\lambda_1,...,\lambda_n\big)$.
Then, the matrices
\be
T_{i,i+1} = \sum_{a,b=1}^n \II^{\otimes(i-1)}\otimes e_{ab} \otimes M^{n_{a} -n_{b}}\otimes \II^{\otimes(N-i-1)}\,,\quad i=1,...,N
\label{formT}
\ee
obey the TL algebra if and only if  
\be
\label{master4}
\forall \{i,j,u\} \subset \{1,\cdots,n\},\quad \left( \sum_{r} \left(\frac{\lambda_j}{\lambda_i}\right)^{n_r} \right) \left(\sum_{k , l} P^{-1}_{i,k} P_{l,j} 
\lambda^{n_k - n_l}_u\right) = n\ \delta_{i,j} .
\ee
\end{lemma}
\prf
Note that the generic gauge covariance of such TL representations
$T_{ij} \rightarrow g_i g_j\, T_{ij}\, g_i^{-1}g_j^{-1}$ allows us to reorder the indices $1,...,n$ in such a way
that $n_a \geq n_b$ when $a \geq b$. Moreover, since only the differences $n_a-n_b$ play a role, up to a global shift, we can 
always assume that $n_a\geq0$, $\forall n_a$.

The form \eqref{formT} automatically solves the one-loop condition \eqref{TLrel0}, so that we only need to
consider the second condition \eqref{TLrel1}. It reads:
\begin{align}
\sum_{i,j,k,l,r} e_{i,j} \otimes M^{n_i - n_r} e_{k,l} M^{n_r - n_j} \otimes M^{n_k - n_l} &= n \sum_{i,j} e_{i,j} \otimes M^{n_i - n_j} \otimes \II
\end{align}
which is equivalent to:
\begin{align}
\sum_{r,k,l} M^{-n_r} e_{k,l} M^{n_r} \otimes M^{n_k-n_l} = n \ \II \otimes \II\,.
\end{align}
We shall now restrict ourselves to matrices $M$ being invertible and diagonalizable. 
Hence we set
$M = P \Lambda P^{-1}$, where $\Lambda$ is an invertible diagonal matrix.

Then the equation becomes
\begin{align}
\sum_{r,k,l} \Lambda^{-n_r} P^{-1} e_{k,l} P \Lambda^{n_r} \otimes \Lambda^{n_k - n_l} &= n \ \II \otimes \II
\end{align} 
or equivalently \eqref{master4} by projecting on $e_{ij}\otimes e_{uu}$.\finprf

\section{Resolution of the TL condition}

We now extract from Eqs. \eqref{master4} the master equations for eigenvalues and eigenvectors of the $M$
matrix. We first need to give some general key definitions for objects which we will come across
in the course of this discussion.

\subsection{Hadamard matrices and master equation}

\begin{defi}
\begin{itemize}
\item A \underline{Complex Hadamard Matrix} (CHM) is an $n\times n$ invertible matrix $U$  obeying
\begin{align}
& |U_{ij}| = 1\,,\quad \forall\ i,j=1,...,n\\
& U = n\, \big(U^{-1}\big)^{\dag}\,,\label{had-dag}
\end{align}
\item A \underline{Generalized complex Hadamard Matrix} (GHM) is an $n\times n$ invertible matrix $U$ with all its entries non-zero and obeying 
the single condition 
\be
 U^{_{-H}} = n\, (U^{-1})^{t}\,,\label{had-inv}
\ee
where $U^{_{-H}}$ is the Hadamard inverse:
$\displaystyle (U^{_{-H}})_{i,j} = \frac1{U_{ij}}$.\\
\item A complex (or generalized) Hadamard matrix $H$ is called \underline{dephased} when all the entries of its first column and first row are equal to one, $H_{1j}=H_{j1}=1$, $\forall\ j$.
\end{itemize}
\label{def:CHM-CGM}
\end{defi}
Remark that the relation \eqref{had-inv} is equivalent to
\be
n\,U_{ij}\,\big(U^{-1}\big)_{ji}=1 \qquad \forall\ i,j=1,...,n
\ee

Real Hadamard matrices (definition \ref{def:CHM-CGM} with real entries $\pm 1$) date back to works of Sylvester \cite{Syl}. 
Complex Hadamard matrices with entries restricted to be roots of unity are also known as Butson matrices, 
introduced in \cite{But}. The situation with generic unimodular entries is described in e.g. \cite{Pol}.
The notion of Hadamard-type criterion for matrices with non-unimodular complex entries, which we introduce in Definition $2$, 
is to the best of our knowledge a new one.

Note that this denomination of ``Generalized Hadamard matrices'' that we have introduced here to denote matrices 
satisfying \eqref{had-inv} must not be confused with the (unfortunately)
similarly-named notion in \cite{But} which involved particular complex Hadamard matrices with an
extra free parameter $k \neq n$: $U^{*} = k\, (U^{-1})^{t}$, and was later dropped to become ``Butson matrices''.

Indeed our object generalizes the notion of a complex Hadamard matrix by replacing the
complex conjugation (an idempotent operation on each matrix element) by the number-inverse,
a similarly idempotent operation naturally extending it to non-unimodular complex numbers.
The transposition operation on the matrix is kept. 
The complex Hadamard condition is then that the inverse of $U$ be given by the transposed of the
complex conjugate matrix \cite{Hadcom}. Any complex Hadamard matrix is therefore 
a generalized Hadamard matrix. The reciprocal problem will be adressed (but not solved) later: can
any GHM be obtained by some well-defined procedure from a CHM ?
\begin{lemma}\label{lem:dephased}
\begin{itemize}
\item If $H$ is a CHM (resp. GHM) then $H'=\sigma_1D_1HD_2\sigma_2$ is also a CHM (resp. GHM), where $D_j$, $j=1,2$ are  unitary (resp. invertible) diagonal matrices and $\sigma_j$, $j=1,2$ are permutation matrices. Two complex (generalized) Hadamard matrices $H$ and $H'$ related in
such a way are called \underline{equivalent}.
\item Any CHM (resp. GHM) is equivalent to a dephased CHM (resp. GHM).
\end{itemize}
\end{lemma}

\medskip

These properties of CHM are to be found in e.g. \cite{Pol}. Their extension to GHM is trivial.

We are now in a position to delve into our issue.
Let us first introduce the matrix $\Omega$  with entries
\be
\label{hadspec}
\Omega_{i,j}= \lambda_i^{n_j}\,,\quad i,j=1,...,n
\ee
hereafter denoted \underline{\textit{Master Matrix}}. 
\begin{prop}
The Master matrix solving \eqref{master4} must be a GHM:
\be
\label{hadom}
 \Omega^{_{-H}} = n\, (\Omega^{-1})^{t}\,.
\ee
Moreover, all the $n_a$'s have to be different, and the spectrum of $M$ must be simple.
\end{prop}
\prf
Equation \eqref{master4} can be rewritten in terms of the matrices $\Omega$ and $P$ as
\begin{equation}\label{eq:master}
 \forall\ i,j,u \;\;\; \left(\Omega^{_{-H}} \Omega^t\right)_{i,j} \left(P^{-1} \Omega^t\right)_{i,u} \left(\Omega^{_{-H}} P \right)_{u,j} = n 
\ \delta_{i,j} 
\end{equation}
Summing equation~\eqref{eq:master} over $i$ or $j$ yields:
\begin{align}
\forall\ i,u, \quad  (P^{-1}\Omega^t)_{i,u} \left(\Omega^{_{-H}} P \,\Omega \,(\Omega^{_{-H}})^t \right)_{u,i} &= n \qquad \text{(summed over }j\text{)}\\
\forall\ j,u, \quad  (\Omega^{_{-H}}P)_{u,j} \left(\Omega \,(\Omega^{_{-H}})^t P^{-1} \Omega^t\right)_{j,u} &= n \qquad \text{(summed over }i\text{)}
\end{align}
Therefore the two matrices $P^{-1} \Omega^t$ and $\Omega^{_{-H}} P$ are full, {\it i.e.} all their entries are non-zero.

It is always consistent to write  $\Omega^{_{-H}} \Omega^t \equiv n \II_n + K$, where $K$ is some matrix with zero diagonal.
Indeed one trivially sees from the definition of $\Omega$ that $\left(\Omega^{_{-H}} \Omega^t\right)_{i,i} = n$ and $K$ therefore measures how far
$\Omega^{_{-H}}$ is from being the matrix inverse (if any) of $\Omega^t$.

From equation \eqref{eq:master} one then gets:
\be
\forall\ i,j,u \;\;\; K_{i,j} \left(P^{-1} \Omega^t\right)_{i,u} \left(\Omega^{_{-H}} P \right)_{u,j} = 0 
\ee
Since we have already established that both matrices $P^{-1} \Omega^t$ and $\Omega^{_{-H}} P$ are full, 
one has necessarily $K_{i,j} = 0$. Hence $\Omega^{_{-H}} \Omega^t = n \II_n$, that is $\Omega$ is invertible and obeys
\eqref{hadom}.

Note immediately that any two integers $n_a$'s have to be distinct otherwise
the matrix $\Omega$ would have at least two identical columns and would not be invertible.

A dual necessary condition is that no two distinctly labeled eigenvalues are equal (which would imply two identical lines in $\Omega$). 
In other words, no degeneracy of eigenvalues is allowed in a realization of the TL condition by diagonalizable $M$ matrices.  \finprf

The TL condition \eqref{master4}, or equivalentely \eqref{eq:master}, therefore factorizes completely into two sets of equations:

-- The one obtained for $i\neq j$ (and trivial at $i=j$) is the polynomial condition expressing that the Master Matrix is a GHM: 
\be
\Omega^{_{-H}} = n (\Omega^{-1})^{t}\quad \mbox{that is} \quad
\sum_{a=1}^{n} \left(\frac{\lambda_i}{\lambda_j}\right)^{n_a} = n\, \delta_{ij}
\label{master2}
\ee
Solving this condition on the Master Matrix will yield simultaneously
consistent sets of powers $n_a$ for $T$ and sets of eigenvalues $\lambda_i$ for $M$.

-- The one obtained for $i=j$ that yields a single condition for the $P$ eigenvector matrices:
\be
\label{master3}
\forall\ i,u, \quad \left(P^{-1} \Omega^t\right)_{i,u} \left(\Omega^{_{-H}} P \right)_{u,i} = 1
\ee
But since $\Omega^{_{-H}}$ is $n$ times the inverse of $\Omega^t$ then $\Omega^{_{-H}} P =n\, (P^{-1} \Omega^t)^{-1}$ 
and therefore \eqref{master3} actually means that the matrix $\Omega^{{-_H}} P$ is a generalized Hadamard matrix 
in the same sense as before (including the $n$ factor). We shall denote it $H$. Hence once the eigenvalues are 
determined by solving the condition \eqref{master2}, the associated consistent $P$ matrices are obtained 
directly from the Master Matrix $\Omega$ once a classification of generalized Hadamard matrices is available. 

\medskip

The problem therefore boils down to two issues, both related to the notion of generalized Hadamard matrices:
\begin{enumerate}
\item Find a classification of the generalized Hadamard matrices $H$ (with complex entries) themselves 
(to get $P$ from $\Omega$ using $H$).
\item Find a characterization and/or a classification of all generalized Hadamard matrices which can be realized 
as Master Matrices, i.e. under the form \eqref{hadspec}, in order to get all consistent sets of $\lambda_i$ 
and $n_a$ obeying \eqref{master2} and the associated master matrix $\Omega$. The integers $n_a$ define a polynomial 
\be
p(z)=\sum_{a=1}^n z^{n_a}
\ee
hereafter called the \underline{\textit{master polynomial}}, and the condition \eqref{master2} expresses that ratios of any two distinct eigenvalues of $M$ are zeroes of $p(z)$. 
\end{enumerate}
From these data one then reconstructs all $M$ matrices as:
\be
M = \Omega^t H \Lambda H^{-1} \Omega^{_{-H}}.
\label{Mform}
\ee

We are now going to partially tackle these two issues.

\subsection{Solving the generalized Hadamard condition}

It must immediately be emphasized at this point that even in the much more studied case of complex Hadamard
matrices no general classification exists. We are thus going to restrict ourselves to a description of the state of 
the art in this case, and a formulation of some exact results allowing to extend it to the generalized Hadamard condition.

Let us now focus on complex Hadamard matrices $(|H_{ij}| = 1)$. A quite complete picture of the current situation
can be found in \cite{BarBen}, see also \cite{Pol}. To give a few salient facts: 
\begin{itemize}
\item[-] The classification is done for $n=2,3,4,5$; 
\item[-] At $n=2,3$
and $5$, only Fourier matrices $\Omega_{ab} \equiv \omega^{(a-1)(b-1)}$ (where $\omega = e^{\frac{2i\pi}{n}}$) 
realize  CHM (up to equivalence);
\item[-] At $n=4$ an extra one-parameter family exists;
\item[-] At $n=6$ several families (including a possibly 
quasi-all-encompassing $4$ parameters family) exist \cite{Szo};
\item[-] Conjectures \cite{BarBen} are available for partial
classifications for $n = p^k$, $p$ prime; $n = a b^k$, $a, b$ prime;
and many individual values of $n$ \cite{Pol}.
\end{itemize}

The first issue now is to try to extend some of these conclusions to dephased generalized Hadamard matrices. Direct solution of the Generalized Hadamard property, by analytic or computer computations, are available for $n=1...4$ and we shall presently give the results of these studies. They unfortunately become  very cumbersome as soon as $n \geq 5$. 

We have however identified a powerful, easily handled tool which generates GHM from CHM (sufficient condition):

{\bf The thickening procedure}\footnote{We borrow this formulation from the notion of ``thickened contours'' 
used by Yu. I. Manin in e.g. Riemann-Hilbert procedures.} consists in starting  from any parametrized 
set $M(a_i)$ of {CHM} such that the complex Hadamard criterion is satisfied \textit{solely}
due to the algebraic relations $a_i \bar a_i = 1$ for all parameters $a_i$. If one substitutes in $M$ the quantity
$\bar a_i$ by $1/a_i$ and relax the conditions $|a_i|=1$,
the resulting set of matrices obeys the generalized Hadamard criterion. This procedure is in particular valid for all families
of parametric complex Hadamard matrices in dimension $4$ and $6$.

As an example let us consider the family $F_4$ of one parameter $n=4$ complex Hadamard matrices. They are
parametrized as:
\begin{equation}\label{ex:F4}
 \Omega = 
 \begin{pmatrix}
  1 & 1 & 1 & 1 \\
  1 & -1 & 1 & -1 \\
  1 & a & -1 & -a \\
  1 & -a & -1 & a
 \end{pmatrix}
\quad \mbox{where } |a|=1.
\end{equation}

If now $a$ is any non-zero complex number, these matrices then become
generalized Hadamard matrices.
 
This procedure may be combined with several classical constructions described hereafter, used for the CHM,
to get many more examples of GHM.

Let us conclude with the cases of dimension $2,3,4$ where we have been able to get a full classification of GHM by explicit
resolution of the equations.
\begin{itemize}
\item[-] at $d=2,3$, GHM are identical to CHM;
\item[-] at  $d=4$, they are all obtained by thickening of CHM. 
\end{itemize}
We have yet no such result at $d=5$, in particular
to get GHM matrices not identical to the Fourier-type CHM (the only such case existing at $d=5$). 

\section{GHM, master matrices and master polynomials}

In this section we explain how to generate larger GHM, with special attention to the construction of master matrices.
Our procedure is based on Di\c{t}\u{a}'s construction of complex Hadamard matrices, which is a generalization of the tensoring procedure.
\subsection{General constructions}
\subsubsection{Fourier matrices}
There exists a general construction that provides one (up to equivalence) CHM which is also a master matrix. The construction can be done in any dimension, and the corresponding matrices are called Fourier matrices.
 
Let $\omega$ be a primitive $n$-th root of unity, {\it i.e.} $\omega=e^{i\ell\frac{2\pi}{n}}$ with $\ell$ prime with $n$. 
The Fourier matrix is defined by 
\be
\Omega_{ab} = \omega^{(a-1)(b-1)},\quad a,b=1,...,n.
\ee
A master matrix being of the form $\lambda_a^{n_b}$, it is natural to identify $\lambda_a = \omega^{a-1}$ and $n_b = b-1$.
Notice that this is not the only solution, for example $n_b = k_b n + b-1$ for some $k_b \in \mathbb{N}$ is also an acceptable identification.

We can then build the master polynomial:
\be
 F_n (z) = \sum_{b=1}^n z^{n_b} = 1 + z + \ldots + z^{n-1} = \frac{z^n - 1}{z-1}.
\ee
The roots of this polynomial are $\frac{\lambda_a}{\lambda_b} = \omega^{a-b}$ for $a \neq b$, as expected.
The solutions proposed in \cite{Chin1,Chin2} belong to this class.

\subsubsection{Di\c{t}\u{a}'s construction}

As for complex Hadamard matrices, if $A$ and $B$ are two generalized Hadamard matrices then $A \otimes B$ is also a generalized Hadamard matrix.
Di\c{t}\u{a} generalized this construction:
\begin{lemma}\label{lemma:Dita}
 Let $A$ be a $n \times n$ complex Hadamard matrix and $\{B^{(1)}, \ldots, B^{(n)}\}$ be a family of $m \times m$ complex Hadamard matrices. 
 Then the $nm \times nm$ matrix:
\be
 C = \begin{pmatrix}
 A_{11} B^{(1)} & A_{12} B^{(1)} & \ldots & A_{1n} B^{(1)} \\
 A_{21} B^{(2)} & A_{22} B^{(2)} & \ldots & A_{2n} B^{(2)} \\
 \vdots & \vdots & \ddots & \vdots \\
 A_{n1} B^{(n)} & A_{n2} B^{(n)} & \ldots & A_{nn} B^{(n)}
 \end{pmatrix}
\ee
is also a complex Hadamard matrix.\\
This statement is also true for generalized Hadamard matrix.
\end{lemma}
The proof can be found in \cite{Dita} for CHM and extends trivially to GHM.

\subsection{Two examples}
Because of lemma \ref{lem:dephased}, we will work with dephased matrices.

\subsubsection{$F_4$ family of complex Hadamard matrices}
The single one-parameter family of complex Hadamard matrices of rank $4$ can be represented by master matrices whenever the parameter $a$ is any root of unity. 
Let $\Omega$ be the matrix given in~\eqref{ex:F4}. It can be associated to the master polynomial $p(z) = (1+z)(1+z^{2k}) = 1 + z + z^{2k} + z^{2k+1}$.
Let
\begin{align}
 \lambda_1 & = 1 &
 \lambda_2 & = -1 &
 \lambda_3 & = e^{\pi i \frac{m}{2k}} = a&
 \lambda_4 & = - e^{\pi i \frac{m}{2k}} = -a
\end{align}
where $m$ is odd.
The master matrix reads $\Omega_{ij} = \lambda_i^{n_j}$, where $n_j$ are the exponents that appear in $p(z)$, {i.e.}
\be
 \{n_1, n_2, n_3, n_4\} = \{ 0,1,2k,2k+1\}
\ee

Notice that varying $m$ and $k$ we get a dense set of $a \in S^1$.

\subsubsection{$F_6$ family of complex Hadamard matrices}

The two-parameter family $F_6$ complex Hadamard matrices of rank $6$ 
\be
 \Omega =
 \begin{pmatrix}
  1 & 1 & 1 & 1 & 1 & 1 \\
  1 & \omega^2 & \omega^4 & 1 & \omega^2 & \omega^4 \\
  1 & \omega^4 & \omega^2 & 1 & \omega^4 & \omega^2 \\
  1 & a & b & -1 & - a & -b \\
  1 & a \omega^2 & b \omega^4 & -1 & -a \omega^2 & -b \omega^4 \\
  1 & a \omega^4 & b \omega^2 & -1 & -a \omega^4 & -b \omega^2
 \end{pmatrix}
\label{F6}
\ee
can be represented 
by master matrices whenever the parameters $a,b$ are both any root of unity.
We remind that in \eqref{F6}, $\omega$ is a 6th root of unity.

We fix three integers $k,r$ and $s$ such that $0<r,s<k$, and consider the polynomial 
\be
p(z) = (1+z^{3r+1}+z^{3s+2})(1+z^{3k}),
\ee
then the exponents $n_i$ are
\be
 \{n_1, n_2, n_3, n_4, n_5, n_6\} = \{ 0,3r+1,3s+2,3k,3k+3r+1,3k+3s+2\}
\ee
We chose the values of $\lambda_i$ to be
\begin{align*}
 \lambda_1 & = 1 &
 \lambda_3 & = \omega^2 &
 \lambda_5 & = \omega^4 \\
 \lambda_2 & = e^{ i \frac{\pi}{3k}} &
 \lambda_4 & = \omega^2 e^{ i \frac{\pi}{3k}} &
 \lambda_6 & = \omega^4 e^{ i \frac{\pi}{3k}}
\end{align*}
It is  easy to check that all ratios ${\lambda_i}/{\lambda_j}$ ($i\neq j$) are roots of $p(z)$.
The master matrix associated to these $\lambda_i$ is exactly \eqref{F6} with $ a= \lambda_2^{3r+1}$ and $b= \lambda_2^{3s+2}$. Varying now $k$, $r$ and $s$ we get a dense set in $S^1 \times S^1$.

In the context of GHM, we allow $a$ and $b$ to be any non-zero complex number.
However we cannot identify the resulting matrix with a master matrix, since for instance one should have $\lambda_4^{n_2 n_4} = a^{n_4} = (-1)^{n_2}$ and therefore $a$ must be a root of unity. 

\subsection{Nesting Fourier matrices}

Both of these examples can be written using Di\c{t}\u{a}'s construction (lemma~\ref{lemma:Dita}). 
For instance, the second one corresponds to the Fourier matrices of size $2 \times 2$ and $3 \times 3$ and a diagonal matrix $D$:
\begin{align}
 A &= \begin{pmatrix}
 1 & 1 \\ 1 & -1
 \end{pmatrix}
&
 B &= \begin{pmatrix}
	1 & 1 & 1 \\
	1 & \omega^2 & \omega^4 \\
	1 & \omega^4 & \omega^2 
 \end{pmatrix}
&
 D = \begin{pmatrix}
 1 & 0 & 0 \\
 0 & a & 0 \\
 0 & 0 & b
 \end{pmatrix}
\end{align}
where we set $B^{(1)} = B$ and $B^{(2)} = B D$.

This process of nesting is already manifest in the way we build the master polynomial.
In what follows we show how to build new solutions nesting smaller solutions, the small block always being Fourier matrices.
This will construct a very large class of solutions.

Let 
\be
 F_{p_1} (z) = \sum_{i=1}^{p_1} z^{g_{1i}\, p_1 + i-1},
\ee
where $g_{1i} \in \mathbb{N}$. Pick the polynomial's root $\omega_1 = e^{\frac{2\pi i}{p_1}}$ and chose $\lambda_i = \omega_1^{f_{1i}\, p_1 + i-1}$.
Then the associated master matrix is the Fourier matrix $\Omega^{(p_1)}_{ij} = \omega_1^{(i-1)(j-1)}$.

We define $F_{p_1 p_2} (z) = F_{p_1} (z) F_{p_2} (z^{\eta_2})$, where $\eta_2 = k_1 p_1$ for some positive integer $k_1$, with the second polynomial being defined in the same way:
\be
 F_{p_2} (z) = \sum_{i=1}^{p_2} z^{g_{2i} p_2 + i-1}.
\ee
.
Let $\omega_2 = e^{\frac{2\pi i}{\eta_2 p_2}}$ and chose 
\be
 \lambda_{i,j} = \omega_1^{f_{1i}\,p_1 + i-1} \omega_2^{f_{2j}\,p_2 + j-1}.
\ee
It is not difficult to show that
\be
 F_{p_1 p_2}\left(\frac{\lambda_{i,j}}{\lambda_{k,\ell}}\right) = n \delta_{ik} \delta_{j\ell},
\ee
where $n = p_1 p_2$.

The master matrix associated to the polynomial $F_{p_1 p_2}$ can be constructed using Di\c{t}\u{a}'s construction: 
\be
 \Omega^{(p_1 p_2)}_{(ij),(k\ell)} = \lambda_{ij}^{(g_{1k}\,p_1+k-1)+\eta_2(g_{2\ell}\,p_2 + \ell-1 )} = \omega_2^{\eta_2 (j-1)(\ell-1)} \Omega_{ik}^{(p_1)} D\left(\omega_2^{f_{2j}\, p_2 + j-1}\right)
\ee
where $D(z)$ is the diagonal matrix:
\be
 \left(D(z)\right)_{k\ell} = \delta_{k\ell} z^{g_{1k}\,p_1 + k-1}.
\ee

This process can now be iterated\footnote{Define $\eta_j = \prod_{i<j} k_i p_i$, where $k_i \in \mathbb{N}$.}, the size of the final matrix being $n= \prod_i p_i$.
In that way, we obtain a large number of examples, including all examples that we were able to construct from known complex Hadamard examples.
An interesting question to tackle would be to understand if this method is complete or to find a counter-example.

Notice that all the entries of the matrix are roots of unity, but the free paramaters $f_{ij}$ and $g_{ij}$ allow us to create a dense set on $S^1$, when varying $k_i$.
Therefore, proving that all examples are obtainable using this method would imply that any master matrix is a CHM, the entries of which are restricted to be roots of unity, i.e. a Butson matrix.

An alternative approach is through the master polynomial.
One can wonder wether  it is possible to find a polynomial $F(z)$ with coefficients in $\{0,1\}$, such that the two following conditions are satisfied: $F(1)=n$ and there is a subset of its roots, $\{\alpha_1, \ldots, \alpha_m\}$, that obeys relations of the type $\alpha_i \alpha_j = \alpha_k$.
Such problems have been studied in \cite{RelPolRoots}, though not exactly in our formulation.

\subsection{Limitations}

There are several limitations of this method.

\begin{itemize}
\item Although it provides a wide spectrum of master matrices and polynomials, we have no proof that it is exhaustive.

\item In the construction of the master polynomials, not all of them correspond to distinct master matrices. For example:
\begin{align*}
 F(z) &= 1 + z^2 + z^3 + z^4 + z^6
\end{align*}
also corresponds to the Fourier matrix based on the root $e^{i\frac{2\pi}{5}}$.

\item Using this construction, we only construct master matrices composed solely by roots of unity.
We must add that \emph{none} of the thickened matrices in $d = 4, 6$ with matrix elements of module different from $1$ are identified as master matrices for any polynomial. For example, if one considers
a matrix of the form \eqref{F6}, only when $a$ and $b$ are module-$1$ complex numbers does $\Omega$ take the form
of a Master Matrix. The same goes if we try to thicken CHM constructed by the above method.
\end{itemize}

\subsection{Non-master Complex Hadamard matrices}

It is important to note that not all complex Hadamard matrices are master matrices.
Two examples:
\begin{align*}
 H_0 &= 
 \begin{pmatrix}
  1 & 1 & 1 & 1 & 1 & 1 \\
  1 & 1 & j & j & j^2 & j^2 \\
  1 & j & 1 & j^2 & j^2 & j \\
  1 & j & j^2 & 1 & j & j^2 \\
  1 & j^2 & j^2 & j & 1 & j \\
  1 & j^2 & j & j^2 & j & 1 
 \end{pmatrix}&
 H_1 &= 
 \begin{pmatrix}
  1 & 1 & 1 & 1 & 1 & 1 \\
  1 & -1 & i & -i & -i & i \\ 
  1 & i & -1 & a & -a & -i \\ 
  1 & -i & -\bar a & -1 & i & \bar a \\ 
  1 & -i & \bar a & i & -1 & -\bar a \\ 
  1 & i & -i & -a & a & -1  
 \end{pmatrix}
\end{align*}
where $j$ is a primitive cubic root of unity, and $a$ is a non-zero complex number.

We prove that $H_0$ is not a Master Matrix. Suppose that $\left(H_0\right)_{ij} = \lambda_i^{n_j}$, where $n_1, \ldots, n_6$ have no common divisor. 
All entries of $H_0$ are a third root of unity, and therefore $\lambda_i$ is a third root of unity.
But there are only three different third roots of unity, which is in contradiction to the fact that $H_0$ has six different rows.

In a similar way we can prove that $H_1$ is not a master matrix either.

\section{Generalized rank-$n$ Ansatz}

We propose finally (and briefly) a generalization of the initial Ansatz. Indeed
the rank-$n$ Ansatz which we started from \eqref{form} can be rewritten in a very illuminating form as:
\be
T = \Big(\sum_{i=1}^{n} e_{ii} \otimes M^{n_i} \Big)\,\Big(\Gamma \otimes \II\Big)\,\Big( \sum_{j=1}^{n} e_{jj}\otimes M^{n_j}\Big)^{-1}
\label{rankn}
\ee
where $\Gamma$ is the particular rank-1 projector $\Gamma \equiv v . v^t$, and $v$ is the  $n$-vector with all components equal to $1$.

Let us now extend this construction to a more general case of rank-1 projector $\Gamma \equiv v . w^t$ where $v$ and $w$ are any two $n$-vectors such that 
$\displaystyle \sum_{i=1}^{n} v_i w_i \equiv \alpha \neq 0$ (i.e. $\Gamma^2=\alpha\,\Gamma$).
Remark that in this construction, one sees immediately that $T$ is of rank $n$:
\be
\rank(T)=\rank(\Gamma \otimes \II)=\rank(\Gamma)\,\rank(\II)=n.
\ee

The TL generators now read, generalizing \eqref{form}:
\be
T_i = \sum_{a,b=1}^n v_a \,w_b \,\II^{\otimes(i-1)}\otimes e_{ab} \otimes M^{n_{a} -n_{b}}\otimes \II^{\otimes(N-i-1)}\,,\quad i=1,...,N
\label{formvw}
\ee
In this generalized situation the whole derivation works out identically to realize representations of the TL algebra 
$TL_N(\sqrt \alpha)$ by the Ansatz \eqref{rankn} at least in the case of diagonalizable $M$ matrices. Keeping
the exact definition of the master matrix $\Omega$ as in \eqref{hadspec} it appears that we must now solve a weighted generalized Hadamard 
condition for $\Omega$
\be
\label{hadomtw}
 \Omega^{_{-H}}V W  = \alpha (\Omega^{-1})^{t}
\ee
Here $V,W$ are Cartan-algebra representations of the vectors $v,w$: $V \equiv \Sigma v_i  e_{ii}$ and $W \equiv \Sigma w_i e_{ii}$.

A quasi-exact (up to replacing $n$ by $\alpha$) Hadamard condition will determine the $P$ matrix but this time for a ``twisted''
 combination involving $V$ and $W$: 
\be
\label{Ptw2}
(P^{-1} V \Omega^t)_{iu} (\Omega^{_{-H}} W P)_{ui} = 1 
\ee
General resolution of the weighted Hadamard condition \eqref{hadomtw} will be left for further studies.

The representation
proposed in \cite{Chin1} takes exactly the form \eqref{rankn} or equivalently \eqref{formvw} albeit with more general vectors $v$, $w$ once the spurious
parameters $q_1,q_2$ are gauged away using the standard gauge covariance for the TL
conditions  $T_{12} \rightarrow g_1g_2 T_{12} (g_1g_2)^{-1}$.

In \cite{Chin1}, after getting rid of the gauge generated by:
\[
 g=
\begin{pmatrix}
q_2 & 0 & 0\\
0 & q_1 & 0\\
0 & 0 & 1
 \end{pmatrix}
\]
one obtains:
\[
 U^{(I)} = 
 \begin{pmatrix}
  1 & 0 & 0 & 0 & \omega & 0 & 0 & 0 & \omega^2\\
  0 & 1 & 0 & 0 & 0 & \omega^2 & \omega & 0 & 0\\
 0 & 0 & 1 & 1 & 0 & 0  & 0 & 1 & 0\\
 0 & 0 & 1 & 1 & 0 & 0  & 0 & 1 & 0\\
  \omega^2 & 0 & 0 & 0 & 1 & 0 & 0 & 0 & \omega\\
  0 & \omega & 0 & 0 & 0 & 1 & \omega^2 & 0 & 0\\
  0 & \omega^2 & 0 & 0 & 0 & \omega & 1 & 0 & 0\\
  0 & 0 & 1 & 1 & 0 & 0  & 0 & 1 & 0\\
  \omega & 0 & 0 & 0 & \omega^2 & 0 & 0 & 0 & 1
 \end{pmatrix}
\]
that takes the form \eqref{formvw} with :
\[
 M=
\begin{pmatrix}
0 & 1 & 0\\
0 & 0 & \omega\\
\omega^2 & 0 & 0
 \end{pmatrix}
\]
In a compact form, $U^{(I)}$ reads:
\be
 U^{(I)} = 
\begin{pmatrix}
M^0 & \omega \, M & \omega \, M^2 \\
\omega^2 \, M^{-1} & M^0 & M\\
\omega^2 \, M^{-2} & M^{-1} & M^0
 \end{pmatrix}
\ee
The extra vectors $v$,$w$  have however the simplifying feature that their associated diagonal matrices obey $V W = 1$ hence the Master Matrix condition \eqref{hadomtw} is not modified. More precisely:
\be
 V=
\begin{pmatrix}
\omega & 0 & 0\\
0 & 1 & 0\\
0 & 0 & 1
 \end{pmatrix} \qquad \mbox{and} \qquad 
W=
\begin{pmatrix}
\omega^2 & 0 & 0\\
0 & 1 & 0\\
0 & 0 & 1
 \end{pmatrix}
\ee
The condition \eqref{Ptw2} associated to $P$  can in this case be rewritten as a non-twisted condition \eqref{master3} for the matrix $\tilde P \equiv V^{-1} P V$. The solutions in \cite{Chin1} are thus very closely related to, but not identical with, matrices $M$ deduced from canonical Fourier-type solutions of the Hadamard conditions.

However due to the degeneracy condition $VW = 1$ this form actually becomes gauge-equivalent in the canonical
TL sense (i.e. $T_{12} \rightarrow g_1g_2\, T_{12}\, (g_1g_2)^{-1}$ ) to the original, pure-power form 
\eqref{formT} with a conjugated $M$ matrix  $\tilde M = g M g^{-1}$. This situation is actually generic: whenever
the diagonal matrices $V$ and $W$, built from the vectors $v$ and $w$, are inverse of each other, the ``general'' rank-$n$ Ansatz with $v$ and $w$ is TL-gauge
equivalent to the standard one.

\section{Conclusion}

We have established an explicit construction of all diagonalizable building blocks $M$ for the Temperley-Lieb representation
Ansatz \eqref{form}. Complex Hadamard matrices and their generalization feature prominently in this construction, both in characterizing
the set of eigenvalues (Master Matrix $\Omega$) and the set of eigenvectors (matrix $P$). It is interesting to remark that the original 
proposition for such generators of TL algebra  \cite{Chin1,Chin2} stemmed from considerations on quantum entanglement: indeed
Complex Hadamard matrices arise in particular in issues related to quantum computation and discrete matrix Fourier transform
(in this last case most specifically Fourier matrices): they define so-called Walsh-Hadamard gates or more general 
quantum gates (see e.g. \cite{KF}). It is thus not a big surprise to see such a connection between TL representations
and Hadamard matrices. 

While eigenvectors are parametrized by GHM, it appears at this stage that all master matrices $\Omega$, encapsulating the eigenvalues of the matrix $M$, constructed explicitly in the previous sections, are complex Hadamard matrices of Butson type (i.e. entries are roots of unity) \cite{But}. It is an open question whether more general master matrices of GHM type may occur; and to determine some sufficient criterion for a GHM to be rewritten as a Master Matrix.

The Butson matrices are the ones that are directly relevant to consideration on quantum entanglement and quantum computations 
issues \cite{Pol}. The GHM however are at this stage not known to have any particular relationship to such problematics.
The issue of their relevance and the relevance of the derived TL representations (with at least eigenvectors described by GHM instead of CHM) to some ``generalized quantum computing'' should be adressed. 

A number of technical issues have been left for further analysis. The most pregnant one is probably the question of non-diagonalizable
(Jordan-like) $M$ matrices. Very preliminary results \cite{GRP} indicate that the notion of master polynomial survives for the
non-degenerate eigenvalues (simple zeroes of the minimal polynomial). The formulation of TL conditions however is much more
complicated due to the occurrence of off-diagonal contributions entangling with the pure eigenvalue-dependent equations.

\subsection*{Acknowledgements}

This work was sponsored by CNRS, Universit\'e de Cergy-Pontoise, Universit\'e de Savoie, and ANR Project DIADEMS (Programme Blanc ANR SIMI1 2010-BLAN-0120-02). PPK is partially supported by 
GDRI ``Formation et recherche en physique th\'eorique'' and RFBR grants 11-01-00570-a, 12-01-00207-a. TF is sponsored
by ANR SIMI1 2010-BLAN-0120-02.

We wish to thank Thierry Huillet for pointing out to us the relevance of the notion of complex Hadamard matrices.


\begin{thebibliography}{99}

\bibitem{TL} H.N.V. Temperley, E. Lieb, \emph{Relations between percolation and colouring problems...}, Proc. Roy. Soc. {\bf A 322} (1971) 251.

\bibitem{Ba} R.J. Baxter, \emph{Exactly Solved Models in Statistical Mechanics}, Academic Press, London (1982).

\bibitem{PM} P. Martin, \emph{Potts models and related problems in Statistical Mechanics}, World Scientific, Singapore (1991).
 
\bibitem{AKR} J. Avan, P.P. Kulish, G. Rollet, \emph{Reflection K-matrices related to Temperley-Lieb R-matrices},
Theor. Math. Phys. {\bf 169(2)} (2011) 1530.

\bibitem{PPK1} P.P. Kulish, \emph{On spin systems related to Temperley-Lieb algebra}, J. Phys. A (Math. Gen.) {\bf 36} (2003) L489.

\bibitem{KMN} P.P. Kulish, N. Manojlovic, Z. Nagy, \emph{Symmetries of spin systems and Birman-Wenzl-Murakami algebra}, J. Math. Phys. {\bf 49} (2008) 023510.

\bibitem{Chin1} G. Wang, T. Hu, C. Zhou, Q. Wang, K. Xue, \emph{Temperley-Lieb algebras, Yang-Baxterization and universal gates}, Quantum Information Processing \textbf{9(6)} (2010) 699, \texttt{arXiv:0903.3711}.

\bibitem{Chin2} C. Sun, G. Wang, T. Hu, C. Zhou, Q. Wang, K. Xue, \emph{The representations of Temperley-Lieb algebras and entanglement in a Yang--Baxter system}, Int. J. Quantum Information \textbf{7} (2009) 1285, \texttt{arXiv:0904.0090}.

\bibitem{Hadcom} 
F. Sz\"{o}ll\H{o}si, \textit{Construction, classification and parametrization of 
complex Hadamard matrices}, PhD thesis, \texttt{arXiv:1110.5590}.

\bibitem{Pol} W. Tadej, K. Zyczkowski, \emph{A concise guide to complex Hadamard matrices},
 Open Systems and Infor. Dyn. \textbf{13} (2006) 133.

\bibitem{CP} V. Chari, A.N. Pressley, \emph{A Guide to Quantum Groups}, Cambridge University Press (1995).

\bibitem{FaTa} L.D. Faddeev, L.M. Takhtadzyan, \emph{The quantum method for the Inverse Problem and the XYZ Heisenberg model}, Usp. Math. Nauk
{\bf 34} (1979) 13.

\bibitem{Jones} V. Jones, \emph{Baxterization}, Int. J. Mod. Phys. \textbf{B4} (1990) 701.

\bibitem{Syl} J.J. Sylvester, \emph{Thoughts on inverse orthogonal matrices, simultaneous sign successions, and 
tessellated pavements in two or more colours, with applications to Newton's rule, 
ornamental tile-work, and the theory of numbers}, Philosophical Magazine \textbf{34} (1867) 461.

\bibitem{But} A.T. Butson, \emph{Generalized Hadamard matrices}, Proc. Am. Math. Soc. {\bf 13} (1962) 894.

\bibitem{BarBen} N. Barros e Sa, I. Bengtsson, \emph{Families of complex Hadamard matrices}, 
Lin. Alg. Appl. \textbf{438} (2013) 2929, \texttt{arXiv:1202.1181}.

\bibitem{Szo} F. Sz\"{o}ll\H{o}si, \emph{Complex Hadamard matrices of order 6: a four-parameter family}, J. London
Math. Soc. {\bf 85} (2012) 616.

\bibitem{Dita} P. Di\c{t}\u{a}, \emph{Some results on the parametrization of complex Hadamard matrices}, J. Phys. A (Math. Gen.) \textbf{37} (2004) 5355.

\bibitem{RelPolRoots} M. Drmota, M. Ska{\l}ba, \emph{Relations between polynomial roots}, Acta Arithmetica \textbf{71} (1995) 65.

\bibitem{KF} K. Fujii, K. Funahashi, T. Kobayashi, \emph{Jarlskog's parametrization of unitary matrices and Qudit theory}, 
 J. Geom. Methods Mod. Phys. \textbf{03} (2006) 269. 

\bibitem{GRP} G. Rollet, private communication.


\end{thebibliography}
\end{document}